\documentclass{cpeauth}

\begin{document}
\def\cop{Copyright \copyright\ 2000 John Wiley \&\ Sons, Ltd.}

\CPE{1}{7}{00}{00}{2000}
\runningheads{K. Keahey et al.} {Fine-Grained Authorizatin in the Grid}

\title{ Fine-Grained Authorization for Job Execution in the Grid: Design and Implementation}

\author{K.~Keahey\footnotemark[2],  V.~Welch\footnotemark[3],  S.~Lang\footnotemark[2],  B.~Liu\footnotemark[4] and  
S.~Meder\footnotemark[3]}

\longaddress{Katarzyna Keahey, Argonne National Laboratory, Mathematics
and Computer Science Division, 9700 S. Cass Ave., Argonne, IL 60439}

\corraddr{Katarzyna Keahey, Argonne National Laboratory, Mathematics
and Computer Science Division, 9700 S. Cass Ave., Argonne, IL 60439}

\footnotetext[2]{Argonne National Laboratory,  Argonne, IL, USA}
\footnotetext[3]{University of Chicago, Chicago, IL, USA}
\footnotetext[4]{University of Houston, Houston, TX, USA}

\cgsn{This work was supported by the Mathematical,
Information, and Computational Sciences Division subprogram of the
Office of Advanced Scientific Computing Research, Office of Science,
SciDAC Program, U.S. Department of Energy, under Contract}{W-31-109-ENG-38.}

\received{1 August 2003}
\revised{1 August 2003}
\noaccepted{}

\begin{abstract}
In this paper we describe our work on enabling fine-grained
authorization for resource usage and management. We address the need
of virtual organizations to enforce their own polices in addition to
those of the resource owners, in regard to both resource consumption
and job management. To implement this design, we propose changes and
extensions to the Globus Toolkit's version 2 resource management
mechanism.  We describe the prototype and the policy language that we
designed to express fine-grained policies, and we present an analysis
of our solution.~\cop
\end{abstract}

\keywords{Grids, Authorization, Policy Enforcement, Resource Management}

\section{INTRODUCTION}

As computational Grids [1] become more widespread, both the resource
pool and the pool of users wishing to use those resources become large
and tend to change dynamically. In such an environment, the
traditional mode of resource sharing, requiring Grid users to
establish direct relationships with resources they wish to use
(i.e. in the form of user accounts), becomes unmanageably complex. We
therefore observe a trend toward defining virtual organizations (VOs)
[1] allowing users to collaborate across different administrative
domains. Credentials issued by such organizations, used in conjunction
with resource provider policies, become the basis of sharing in
Grids. In this model, resource providers typically outsource some
subset of their policy administration to the VO. This strategy allows
the VO to coordinate policy across resources in different domains
forming a consistent policy environment in which its participants can
operate. Such an environment requires mechanisms for enabling the VO
to specify and enforce VO-specific policies on tasks and resources owned by VO
participants.

Another developing trend is the need to express and enforce fine-grain
policies on the usage of resources and services. These can no longer
be expressed by simple access control; resource owners and VO
administrators may want to specify exactly what fractions or
configurations of resource may be used by a given entity. In addition,
while some VOs are focused on sharing of hardware resources (e.g.,
CPUs and storage), for others the primary motivation is to coordinate
sharing of application services [2] requiring access to both software
and hardware. In these cases the VO members should not be running
arbitrary code but only applications sanctioned by VO policy. Such
policies may be dynamic, adapting over time or even changing during
application execution, depending on factors such as past and current
resource utilization record, a member's role in the VO, deadline-based
priorities.

In this paper, we address the requirements posed by these two
trends. We present a design for service and resource management that
enables a VO and resource managers to specify fine-grained service and
resource usage policies using VO credentials and allows resources to
enforce those policies. We implement our design as extensions to the
Globus Toolkit version 2 (GT2) resource management mechanism [3]. We
then consider policy enforcement in the context of two types of policy
target: application services and traditional computing resources. A
prototype of this implementation, combined with the Akenti
authorization system [4], was demonstrated at the SC02 conference and
is currently being adopted by the National Fusion Collaboratory [2].

This paper is organized as follows. In Section 2, we present a use
case scenario and concrete requirements guiding our design. In Section
3 we define our problem. We follow this by a discussion of the
capabilities of the Globus Toolkit's resource management (GRAM) [3]
mechanism (Section 4) and describe extensions needed to GRAM to support our
architecture (Section 5).  In the last three sections, we analyze our solution, present future
directions, and conclude the paper.

\section{USE CASE SCENARIO AND REQUIREMENTS}

In a typical VO scenario, a resource provider has reached an agreement
with a VO to allow the VO to use some resource allocation. The
resource provider thinks of the allocation in a coarse-grained manner:
the provider is concerned about how many resources the VO can use as a
whole, not about how allocation is used inside the VO.

The finer-grained specification of resource usage among the VO
participants is the responsibility of the VO. For example, the VO has
two primary classifications of its members:

\begin{enumerate}

\item [$\bullet$] One group is developing, installing, and debugging the
application services used by the VO to perform a scientific
computation. This group may need to run many types of processes
(e.g., compilers, debuggers, applications services) in order to debug
and deploy the VO application services, but should be consuming small
amounts of traditional computing resources (e.g., CPU, disk and
bandwidth) in doing so.

\item [$\bullet$] The second group performs analysis using the
application services. This group may need to consume large
amounts of resources in order to run simulations related to their research.
\end{enumerate}

Thus, the VO may wish to specify finer-grained policies that allow certain
users to use more or fewer resources than other users. These policies may
be dynamic and change at any point (for example, during runtime of an
application).

In addition to policy on resource utilization, the VO wishes to be
able to manage jobs running on VO resources. For example, users often
have long-running computational jobs using VO resources, which the VO
often has short-notice high-priority jobs that require immediate
access to resources. This mode of operation requires suspending
existing jobs to free up resources, something that normally only the
user that submitted the job has the right to do. Since going through
the user who submitted the original job may not always be an option,
the VO wants to give a group of its members the ability to manage any
jobs using VO resources so they can instantiate high-priority jobs on
short notice.

Supporting this scenario places several requirements on the authorization policy system:

\begin{enumerate}

\item {\em Combining policies from different sources}. In outsourcing
a portion of the policy administration to the VO, the policy enforcement
mechanism on the resource needs to be able to combine policies from
two different sources: the resource owner and the VO.

\item {\em Fine-grained control of how resources are used}. For the VO
to express the differences between how its user groups are allowed to
use resources, the VO needs to be able to express policies regarding a
variety of aspects of resource usage, not just grant access.

\item {\em VO-wide management of jobs and resource allocations}. The
VO wants to be able to treat jobs as resources themselves that can be
managed. This requirement poses a particular challenge because jobs are dynamic, so
static methods of policy management are not effective. Users may also
start jobs that shouldn't be under the domain of the VO; for example, a user
may have allocations on a resource other than those obtained through
the VO, and jobs invoked under this alternate allocation should not be
subject to VO policy.

\item {\em Fine-grained, dynamic enforcement mechanisms}. In order to
support any policies, there must be enforcement mechanisms capable of
implementing these policies. Most resources today are capable of
policy enforcement at the user level: that is, all jobs run by a given
user will have the same policy applied to them. These mechanisms are
typically statically configured through file permissions, quotas and
similar mechanisms. Our scenario brings out the requirement that
enforcement mechanisms need to handle dynamic, fine-grained policies.
\end{enumerate}

\section{INTERACTION MODEL}

To support the scenario described in the preceding section, we need to
provide resource management mechanisms that allow the specification
and consistent enforcement of authorization and usage policies that
come from both the VO and the resource owner. In addition to allowing
the VO to specify policies on standard computational resources, such as
processor time and storage, we need to allow the VO to specify
policies on application services that it deploys, as well as
long-running computational jobs instantiated by VO members.

In our work we assume the following interaction model:
\begin{enumerate}
\item A user submits a request, composed of the job's description to initiate a job. The request is accompanied by the user's Grid credentials, which may include the user's personal credentials as well as VO-issued credentials. 
\item This request is evaluated against both local and VO policies by different policy evaluation points (PEPs), capable of interpreting the VO and the resource management policy respectively, located in the resource management facilities.  
\item If the request is authorized by both PEPs, it is mapped to a set of local resource credentials (e.g., a Unix user account). Policy enforcement is carried out by local enforcement mechanisms operating based on local credentials. 
\item During the job execution, a VO user may make management requests to the job (e.g., request information, suspend or resume a job, cancel a job). 
\end{enumerate}

\section{GRID RESOURCE MANAGEMENT IN GT2}

The Globus Toolkit provides mechanisms for security, data management
and movement, resource monitoring and discovery (MDS) and resource
acquisition and management. In this paper we are focusing on the
functionality of resource acquisition and management, which is
implemented by the GRAM (Grid Resource Acquisition and Management)
system [3].

The GRAM system has two major software components: the Gatekeeper and
the Job Manager. The Gatekeeper is responsible for translating Grid
credentials to local credentials (e.g., mapping the user to a local
account based on their Grid credentials) and creating a Job Manager
Instance to handle the specific job invocation request. The Job
Manager Instance (JMI) is a Grid service that instantiates and then
provides for the ability to manage a job.  Figure 1 shows the
interaction of these elements; in this section we explain their roles
and limitations.

\subsection{Gatekeeper}

The Gatekeeper is responsible for authenticating the requesting Grid
user, authorizing their job invocation request and determining the
account in which their job should be run. Authentication, performed by
the Grid Security Infrastructure [5], verifies the validity of the
presented Grid credentials, the user's possession of those
credentials, and the user's Grid identity as indicated by those
credentials. Authorization is based on the user's Grid identity and a
policy contained in a configuration file, the gridmapfile, which
serves as an access control list. Mapping from the Grid identity to a
local account is also done with the policy in the gridmapfile,
effectively translating the user's Grid credential into a local user
credential. Finally, the Gatekeeper starts up a Job Manage Instance,
executing with the user's local credential. This mode of operation
requires the user to have an account on the resource and implements
enforcement by privileges of the account.

\begin{figure}
\centering\includegraphics{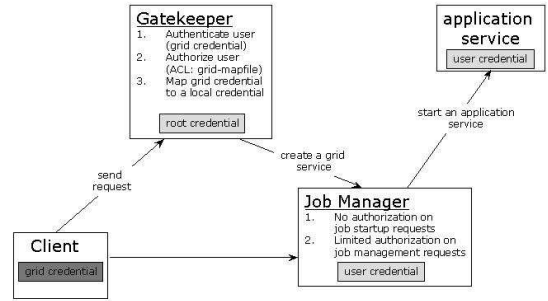}
\caption{Interaction of the main components of GRAM}
\end{figure}

\subsection{Job Manager Instance}

The JMI parses the user's request, including the job description, and
interfaces with the resource's job control system (e.g., LSF, PBS) to
initiate the user's job. During the job's execution the JMI monitors
its progress and handles job management requests (e.g., suspend,
stop, query) from the user. Since the JMI is run under the user's
local credential, as defined by the user's account, the operating
system and local job control system are able to enforce local policy
on the JMI and user job by the policy tied to that account.

The JMI has no authorization on job startup since the Gatekeeper has
already authorized it. Once the job has been started however, the JMI
accepts, authenticates, and authorizes management requests on the
job. In GT2, the authorization policy on these management requests is
static and simple: the Grid identity of the user making the request
must match the Grid identity of the user who initiated the job.

\subsection{GRAM Shortcomings}

The current GRAM architecture has a number of shortcomings when matched against the requirements we laid out in Section 2:

\begin{enumerate}
\item Authorization of user job startup is coarse-grained. It is based solely on whether a user has an account on a resource.
\item Authorization on job management is coarse-grained and static. Only the user who initiated a job is allowed to manage it. 
\item Enforcement is implemented chiefly through the medium of privileges tied to a statically configured local account (JMI runs under local user credential) and is therefore useless for enforcing fine-grained policy or dynamic policy coming from sources external to the resource (such as a VO).
\item Local enforcement depends on the rights attached to the user's account, not the rights presented by the user with a specific request; in other words, the enforcement vehicle is largely accidental.
\item A local account must exist for a user; as described in the introduction, this creates an undue burden on system administrators and users alike. This burden prevents wide adoption of the network services model in large and dynamically changing communities.
\end{enumerate}

These problems can, and have been, in some measure alleviated by
clever setup. For example, the impact of (4) can be alleviated by
mapping a grid identity to several different local accounts with
different capabilities. Often, (5) is handled by working with
``shared accounts'' (which, however, introduce many security, audit,
accounting and other problems) or by providing a limited
implementation of dynamic accounts [6,13,14].

\section{AUTHORIZATION AND ENFORCEMENT EXTENSIONS TO GRAM}

In this section we describe extensions to the GT2 Grid Resource
Acquisition and Management (GRAM) that address the shortcomings
described above.

\begin{figure}
\centering\includegraphics{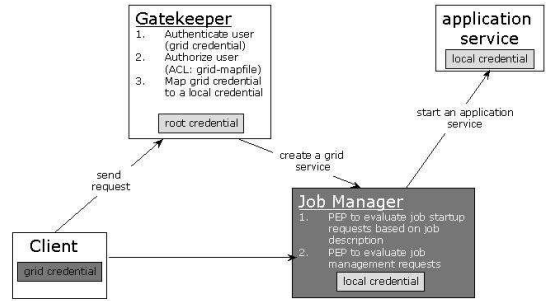}
\caption{Changes to GRAM: the changed component (the Job Manager) has been highlighted in gray}
\end{figure}

We extended the GRAM design to allow authorization callouts,
 evaluating the user's job invocation and management requests in the
 context of policies defined by the resource owner and VO. Our changes
 to GRAM, prototyped using GT2, are illustrated in Figure 2. In our
 prototype we experimented with policies written in plain text files
 on the resource. These files included both local resource and VO
 policies (in a real system the VO policies would be carried in the VO
 credentials).  This work has recently been tested with the Akenti [4]
 system, representing the same policies as described here, and is being
 adopted by the National Fusion Collaboratory [2]. In order to show
the generality of our approach, we also experimented with the
 Community Authorization Service (CAS) [7]. Both of these systems
 allow for multiple policies sources but have significant
 differences, in terms of both architecture and programming APIs.

\subsection{Policy Language}

GRAM allows users to start and manage jobs by submitting requests
composed of an action, (e.g., initiate, cancel, provide status, change
priority) and, in the case of job initiation, a job
description. The job description is formulated in terms of attributes
using the Resource Specification Language (RSL) [3]. RSL consists of
attribute value pairs specifying job parameters referring to
executable description (executable name, directory where it is
located, etc.) and resource requirements (number of CPUs to be used,
maximum/minimum allowable memory, maximum time a job is allowed to
run, etc.).

We have designed a simple policy language that allows for policy
specification in terms of RSL. The policy assumes that unless a
specific stipulation has been made, an action will not be
allowed. Otherwise, a user, or a group of users, is related to a set
of assertions. The rules have the form of user (or group) identity
separated by a colon from a set of action based assertions that follow
the RSL syntax.

To express the rules, we extended the RSL set of attributes with the addition of the following:

\begin{enumerate}
\item [$\bullet$] {\em Action.} This attribute represents what the
user wants to do with the job. Currently, it can take values of
``start'', ``cancel'', ``information'', or ``signal'', where
``signal'' describes a variety of job management actions such as
changing priority.

\item [$\bullet$] {\em Jobowner.} The jobowner attribute denotes the
job initiator and can take values of the distinguished name of a
job initiator's Grid credential. It is used mainly to express VO-wide
management policy.

\item [$\bullet$] {\em Jobtag.} The jobtag attribute has been
introduced in order to enable the specification of VO-wide job
management policies.  A jobtag indicates the job membership in a group
of jobs for which policy can be defined. For example, a set of users
with an administrative role in the VO can be granted the right to
manage all jobs in a particular group. A policy may require a VO user
to submit a job with a specific jobtag, hence placing it into a group
that is manageable by another user (or group of users). At present,
jobtags are statically defined by a policy administrator.
\end{enumerate}

We also added the following values to RSL:
\begin{enumerate}
\item [$\bullet$] ''NULL'' to denote a nonempty value

\item [$\bullet$] ''SELF'' to allow expression of the job initiator's identity in a policy.
\end{enumerate}

These extensions allow the following types of assertions to be expressed in policy:
\begin{enumerate}

\item [$\bullet$] The job request is permitted to contain a particular
attribute, value, or set of values. This extension allows one, for
example, to limit the maximum number of processors used or to restrict
the name of the executable to a specified set. Multiple assertions can
be made about the same attribute.

\item [$\bullet$] The job request is required to contain a particular attribute, possibly with a particular value or set of values. For example, the job request must specify a jobtag attribute to allow its management by a VO-defined group of administrators.

\item [$\bullet$] The job request is required not to contain a
particular attribute. For example, the job request must not specify a
particular queue, which is reserved for high-priority users.
\end{enumerate}

Our extensions allow a policy not only to limit the usage of
traditional computational resources but also to dictate the
executables they are allowed to invoke, allowing a VO to limit
resource consumption. Further, by introducing the notion of a jobtag,
we are able to express policies allowing users to manage jobs. The
example below illustrates how policy may be expressed.

\begin{verbatim}
&/O=Grid/O=Globus/OU=mcs.anl.gov: 
(action = start)(jobtag != NULL)

/O=Grid/O=Globus/OU=mcs.anl.gov/CN= Bo Liu:
&(action = start)(executable = test1)(directory = /sandbox/test)(jobtag = ADS)(count<4)
&(action = start)(executable = test2)(directory = /sandbox/test)(jobtag = NFC)(count<4)

/O=Grid/O=GlobusOU=mcs.anl.gov/CN= KateKeahey:
&(action = start)(executable = TRANSP)(directory = /sandbox/test)(jobtag = NFC)
&(action=cancel)(jobtag=NFC)

\end{verbatim}

The first statement in the policy specifies a requirement for a group
of users whose Grid identities start with the string {\tt ``
/O=Grid/O=Globus/OU=mcs.anl.gov''}. The requirement is that for job
invocations (where the action is ``start''), the job description must
contain a jobtag attribute with some value. This allows us to later
write management policies referring to that jobtag.  

The second statement in the policy refers to a specific user, Bo Liu,
and states that she can start jobs only using the ``test1'' and
``test2'' executables. The rules also place constraints on the
directory from which the executable can be taken and the jobtag they
can be started with. In addition, a constraint is placed on the number
of processors Bo Liu can use ($count < 4$).

The third statement in the policy gives
user Kate Keahey the right to start jobs using the ``TRANSP''
executable from a specific directory and with a specific jobtag. It
also gives her the right to cancel all the jobs with jobtag ``NFC'',
for example, jobs based on the executable ``test1'' started by Bo Liu.

\subsection{Enforcing Policies in GRAM}

We enforce our policies in GRAM by creating a policy evaluation point
(PEP) controlling all external access to a resource via GRAM; an
action is authorized depending on decision yielded by the PEP. Policy
can be enforced in GRAM at multiple PEPs corresponding to different
decision domains; for example, a PEP placed in the Gatekeeper can allow
or disallow access based on the user's Grid identity. Since our work
focuses on job and resource management, we established a PEP in the Job
Manager (JM). The JM parses user job descriptions and can therefore
evaluate policy that depends on the nature of the job request in
addition to the user's identity.

Specifically, our additions consist of the following:

\begin{enumerate}

\item [$\bullet$] {\em An authorization callout API to integrate the PEP
with the JM}. The callout passes to the PEP authorization module the
relevant information, such as the credential of the user requesting a
remote job, the credential of the user who originally started the job,
the action to be performed (such as start or cancel a job), a unique
job identifier, and the job description expressed in RSL. The PEP
responds through the callout API with either success or an appropriate
authorization error. This call is made whenever an action needs to be
authorized, that is, before creating a job manager request and before
calls to cancel, query, and signal a running job.

\item [$\bullet$] {\em Policy-based authorization for job
management}. As discussed in Section 4, each job management request
other than job startup is currently authorized by GRAM so that only
the user that started a job is allowed to manage it. We modified the
authorization in GRAM to enable Grid users other than the job
initiator to manage the job based on policy with decisions rendered
through the authorization callout API. In addition to changes to the
authorization model, this modification also required extensions to the
GRAM client allowing the client to process other identities than that
of the client (specifically, allowing it to recognize the identity of
the job originator).

\item [$\bullet$] {\em RSL parameters}. We extended RSL to add the ``jobtag'' parameter allowing the user to submit a job to a specific job management group.

\item [$\bullet$] {\em Errors}. We further extended the GRAM protocol to return authorization errors describing reasons for authorization denial as well as authorization system failures.
\end{enumerate}

For easy integration of third-party authorization
solutions, the callout API provides facilities for runtime
configurable callouts.  Callouts can be configured either through a
configuration file or an API call. Configuration consists of
specifying an abstract callout name, the path to the dynamic library
that implements the callout, and the symbol for the callout in the
library. Callouts are invoked through runtime loading of dynamic
libraries using GNU Libtool's dlopen-like portability library.
Arguments to the callout are passed by using the C variable argument list
facility.

The insertion of callout points into JM required defining a GRAM
authorization callout type, (i.e., an abstract callout type), the exact
arguments passed to the callout, and a set of errors the callout may
return.  These callout points are configured by parsing a global
configuration file.

\section{ANALYSIS}

Our solution overcame some of the shortcomings outlined in Section
4.3. However our approach has a number of outstanding issues that we
discuss in this section.

\subsection{Gateway Enforcement Model}

A weakness of the gateway approach is that once a gateway authorizes
an action (for example, a job execution) it is no longer involved in
the continuous enforcement of the policy. GRAM maps incoming
requests to static local accounts to perform this continuous policy
enforcement.

This has two consequences: (1) the local policy enforcement depends
on the privileges tied to the account that the user maps to on the local
system, rather than to the credential with which the request was made,
and (2) GRAM's abilities for continuous policy enforcement are limited
by local capabilities for policy enforcement.

The first limitation could to some extent, be dealt with by using
dynamic accounts [6,13,14]. Dynamic accounts are accounts created and
configured on the fly by a resource management facility. This enables
the resource management system to run jobs on a system for users that
do not have an account on that system, and it also enables account
configuration relevant to policies for a particular resource
management request as opposed to a static user's configuration. To
some extent a dynamic account can be also used as a sandbox on the
user's rights (by modifying user's group membership to control file
system access, for example). Although work has been
done to support fine-grained policy for file access [8], Unix
accounts allow the user to modify only very few configuration
parameters, and hence the enforcement implemented in an account is
coarse-grained. 

A sandbox is an environment that imposes restrictions on resource
usage [9,15,16]. Sandboxing represents a strong enforcement solution, having
the resource operating system act as the policy evaluation and
enforcement modules, and is complementary to the gateway
approach. However, while the sandboxes provide a solution with relatively high
degree of security, they are hard to implement portably and may
introduce a performance penalty.

\subsection{Job Manager Trust Model}

In the GRAM architecture, the job manager runs with the user's local
credentials; this approach makes the job manager less than ideal for
policy enforcement. The reasons are twofold. First, from the security
perspective it is vulnerable to user tampering that could result in
changes in policy enforcement. Second, it effectively limits
enforcement potential for VO-wide job management. For example, a user
managing a job may cancel a job started by somebody else (by virtue of
the fact that the job manager is running with the job initiator's
local credential), but the user may not apply higher resource rights
to, for example, raise the job's priority.

One possible solution to this problem in the context of the GRAM
architecture would be to locate the policy enforcement point in the
gatekeeper. However, this would increase the vulnerability of the
system by placing more complex code into the trusted component of the
system, increasing chances for logic errors, buffer overflows, and so
forth.

Another possibility would be for policy enforcement to be done by
trusted services such as the local operating system. As discussed
earlier, this is difficult today because most operating systems do not
have the support for fine-grained policies that we
require. Investigation into sandboxing techniques remains an open
research issue.

\subsection{Policy Language}

Our implementation currently expresses policy in terms of the same
resource specification language (RSL) that GRAM uses to describe
jobs. While this allows for easy comparison of a job description with
a policy, it is not a standard policy language. Policy administrators
are not familiar with RSL, and our initial experiences show that
expressing policies in these terms is not natural to this
community. This difficulty is compounded by the fact that the syntax
is not be supported by standard policy tools. We are therefore
investigating existing policy languages as a replacement to our
RSL-based scheme. With the merging of Grid technologies and Web
service-based technologies in OGSA[10], languages based on XML, such
as XACML [11] and XrML [12], are being scrutinized by the Grid
security community in general and are viable candidates.

\subsection{Relevance to Other Systems}

Our work could be applied to systems similar to the Globus Toolkit
based on its relevance. For example, Legion authorization is
implemented by the use of a MayI [20] method on all Legion objects. In
the default implementation, this method offers similar functionality
as the Globus Toolkit, with access control lists and static mapping to
local accounts. Our work could be integrated with Legion in a similar
manner as we described here, through the reimplementation of object
creation routines (Legion's equivalent of GRAM) and the MayI method.
Condor's [21] interface on the other hand, is based more on compute
resources than instantiated jobs. Although it also uses access
control lists to manage its policy, it does not provide the per-job
interface of GRAM and Legion. This makes our work less relevant to that system.

\section{TOWARD GT3: FUTURE DIRECTIONS}

To address the open issues summarized above, we are developing an
architecture building on abstractions and mechanisms defined as part
of the Open Grid Services Infrastructure (OGSI) [17]. The key to the
policy enforcement questions is the implementation of an abstraction
that would allow for dynamic creation and management of a local
protection environment (such as a Unix account, a sandbox [9,15,16], or a
virtual machine [18, 19]). Such abstraction would not only provide
protection but also facilitate resource management (by enforcing
limits on resource usage for a particular user) and maintain state
associated with its owner. We will call such an abstraction a {\em
dynamic session}.

The OGSI abstractions of Grid Service and Grid Service factory are
suitable for this task of implementing such abstraction. Representing
a session as a Gird service will provide uniform management
capabilities across different technologies that could be used to
implement sessions. Standardizing session creation alleviates the
administrative burden involved in adding users to a VO, and it also
allows session creation based on rights granted to a particular user
at a specific time. To manage dynamic sessions, we can
leverage the OGSI Service Data Element (SDE) mechanism in order to make the
properties of a session (such as its termination time) accessible to
the session owner and modifiable by him or her.

In a typical interaction, a user requests a session with certain
properties (i.e., resource requirements) from a session factory. The
factory authorizes the request and, on success, creates dynamic
session service and a local protection environment corresponding to
it. As part of the creation process, policy defining sharing rights
for the session is written. This policy can be modified by authorized
entities during the service's lifetime, as can other session
properties. The user can submit against that session, pending
conformance with the rights just created. Further, to facilitate
management of sessions that do not have to be reused multiple times
(i.e., do not preserve state between the times when they get used), the
resource management service can obtain sessions based on credentials
presented by the user requesting job submission or credentials of the
resource manager itself.

\section{CONCLUSIONS}

We have described the design and implementation of an authorization
system allowing for enforcement of fine-grained policies and VO-wide
management of remote jobs. To implement this design, we have proposed
changes to the Globus Toolkit GRAM design and have designed a policy
language suitable for our needs. We are planning to use the same
mechanism to provide pluggable authorization in other components of
the Globus Toolkit.

\acks

We are pleased to acknowledge contributions to this work by Mary
Thompson of LBNL. 

\section{\bf REFERENCES}

\begin{enumerate}

\item Foster, I., C. Kesselman, and S. Tuecke, The Anatomy of the
Grid: Enabling Scalable Virtual Organizations. International Journal
of High Performance Computing Applications, 2001. 15(3): p. 200-222.

\item Keahey, K., T. Fredian, Q. Peng, D.P. Schissel, M. Thompson,
I. Foster, M. Greenwald, and D. McCune, Computational Grids in
Action: the National Fusion Collaboratory. Future Generation
Computing Systems (to ap-pear), October 2002. 18(8): p. 1005-1015.

\item Czajkowski, K., I. Foster, N. Karonis, C. Kesselman, S. Martin,
W. Smith, and S. Tuecke, A Resource Management Architecture for
Meta-computing Systems, in 4th Workshop on Job Scheduling Strategies
for Parallel Processing. 1998, Springer-Verlag. p. 62-82.

\item Thompson, M., W. Johnston, S. Mudumbai, G. Hoo, K. Jackson, and
A. Essiari, Certificate-based Access Control for Widely Distributed
Resources, in Proc. 8th Usenix Security Symposium. 1999.

\item Butler, R., D. Engert, I. Foster, C. Kesselman, S. Tuecke,
J. Volmer, and V. Welch, Design and Deployment of a National-Scale
Authentication Infrastructure. IEEE Computer, 2000. 33(12): p. 60-66.

\item Dynamic Accounts. http://www.gridpp.ac.uk/gridmapdir/.

\item Pearlman, L., V. Welch, I. Foster, C. Kesselman, and
S. Tuecke, A Community Authorization Service for Group
Collaboration. in IEEE Workshop on Policies for Distributed Systems
and Networks. 2002.

\item Lorch M. and D. Kafura, Supporting Secure Ad-hoc User
Collaboration in Grid Environments. in Proceedings of the 3rd
Int. Workshop on Grid Computing - Grid 2002, Baltimore, MD, USA. 2002.

\item Chang, F., A. Itzkovitz, and V. Karamacheti, User-level
Resource-constrained Sandboxing. Proceedings of the USENIX Windows
Systems Symposium (previously USENIX-NT), 2000.

\item Foster, I., C. Kesselman, J. Nick, and S. Tuecke, The Physiology
of the Grid: An Open Grid Services Architecture for Distributed
Sys-tems Integration. Open Grid Service Infrastructure WG,
Global Grid Forum, 2002.

\item OASIS eXtensible Access Control Markup Language (XACML)
Committee Specification 1.0 (Revision
1). http://www.oasis-open.org/committees/xacml/docs/s-xacml-specification-1.0-1.doc,
2002.

\item XRML. http://www.xrml.org/get\_XrML.asp.
\item Hacker, T. and B. Athey, A Methodology for Account Management in Grid Computing Environments. Proceedings of the 2nd International Workshop on Grid Computing, 2001.
\item Kapadia, N. H., R. J. Figueiredo, and J. Fortes. Enhancing the Scalability and Usability of Computational Grids via Logical User Accounts and Virtual File Systems. in 10th Heterogeneous Computing Workshop. 2001. San Francisco, California.
\item Bosilca, G., F.  Capello, A. Djilali, G. Fedak, T. Hernault and F. Magniette, Performance Evaluation of Sandboxing Techniqes for Peer-to-Peer Computing. 
\item Goldberg, I., D. Wagner, R. Thomas, and E. Brewer, A Secure Environment for Untrusted Helper Applications --- Confining the   Wily Hacker, in Proc. 1996 USENIX Security Symposium. 1996.
\item Tuecke, S., K. Czajkowski, I. Foster, J. Frey, S. Graham, and C. Kesselman, Grid Service Specification. 2003: Open Grid Service Infrastructure WG, Global Grid Forum.
\item VMware: http://www.vmware.com/.
\item User Mode Linux (UML). http://user-mode-linux.sourceforge.net/.

\item Humphrey, M., F. Knabe, A. Ferrari and A. Grimshaw,
Accountability and Control of Process Creation in Metasystems. 2000
Network and Distributed System Security Symposium, 2000.

\item ``Condor Version 6.4.7 Manual: Security In Condor'', \\
http://www.cs.wisc.edu/condor/manual/v6.4/3\_7Security\_In.html, 2003.

\end{enumerate}

\end{document}